\documentclass[preprint]{aastex62}

\newcommand\asec{$''$}
\newcommand\ha{H$_{\alpha}$}
\newcommand\kms{~km~s$^{-1}$}

\usepackage{natbib}

\received{December 5, 2019}
\revised{January 20, 2020}
\accepted{February 23, 2020}
\submitjournal{ApJ Letters}

\shorttitle{Rapid Evolution of Type II Spicules}
\shortauthors{Yurchyshyn et al.}

\begin{document}

\title{Rapid Evolution of Type II Spicules Observed in Goode Solar Telescope On-Disk \ha\ Images}

\correspondingauthor{Vasyl Yurchyshyn}
\email{vasyl.yurchyshyn@njit.edu}

\author[0000-0001-9982-2175]{Vasyl Yurchyshyn}
\affiliation{Big Bear Solar Observatory, New Jersey Institute of Technology, \\
40386 North Shore Lane, Big Bear City, CA 92314, USA}
\author[0000-0003-2427-6047]{Wenda Cao}
\affiliation{Big Bear Solar Observatory, New Jersey Institute of Technology, \\
40386 North Shore Lane, Big Bear City, CA 92314, USA}
\author[0000-0001-6466-4226]{Valentina Abramenko}
\affiliation{Crimean Astrophysical Observatory of Russian Academy of Science, Nauchny, Bakhchisaray, Russia}
\author[0000-0002-3238-0779]{Xu Yang}
\affiliation{Big Bear Solar Observatory, New Jersey Institute of Technology, \\
40386 North Shore Lane, Big Bear City, CA 92314, USA}
\author[0000-0003-2161-9606]{Kyung-Suk Cho}
\affiliation{Space Science Division, Korea Astronomy and Space Science Institute, Daejeon 305-348, Republic of Korea}
\affiliation{Department of Astronomy and Space Science, University of Science and Technology, \\
Daejeon 305-348, Republic of Korea}

\begin{abstract}

We analyze ground-based chromospheric data acquired at a high temporal cadence of 2~s in wings of the \ha\ spectral line using Goode Solar Telescope (GST) operating at the Big Bear Solar Observatory. We inspected a 30 minute long \ha-0.08~nm data set to find that rapid blue-shifted \ha\ excursions (RBEs), which are a cool component of type II spicules, experience very rapid morphological changes on the time scales of the order of 1 second. Unlike typical reconnection jets, RBEs very frequently appear \textit{in situ} without any clear evidence of \ha\ material being injected from below. Their evolution includes inverted ``Y'', ``V'', ``N'', and parallel splitting (doubling) patterns as well as sudden formation of a diffuse region followed by branching. We also find that the same feature may undergo several splitting episodes within about 1 min time interval.

\end{abstract}


\section{Introduction} \label{sec:intro}

Large- and small-scale jets and upflows observed in the lower atmosphere of quiet Sun (QS) areas are considered to play an important role in the transfer of mass and energy from the dense chromosphere into the corona. However, their origin and connection to the dynamics of the magnetic fields are not well understood and explored. 

Type  II spicules were first discovered in off-limb \textit{Hinode} data \citep{2007PASJ...59S.655D}. They are short-lived ($<$ 100~s), thin ($<$ 0\asec.7), structures seen everywhere in \textit{Hinode} Ca II images that show high Doppler velocities \citep[50-150\kms,][]{2007PASJ...59S.655D} and return flows \citep{2014ApJ...792L..15P}. When observed in IRIS data, they show higher apparent speeds of 80-300\kms\ \citep{2014Sci...346A.315T,2016SoPh..291.1129N}. Type II spicules are omnipresent and they carry a large amount of magnetic energy \citep{2007Sci...318.1574D,2011Natur.475..477M,torsional_osc,Liu2019}.

Type II spicules were found to have on-disk counterparts. They have been identified with Ca II ``straws'' and rapid blue-shifted excursions (RBEs) seen in Ca II 854.2~nm \citep{2008ApJ...679L.167L} and \ha\ lines on the solar disk \citep[e.g.,][]{2006ASPC..354..276R,2009ApJ...705..272R,2015ApJ...802...26K}. Here we use the term type II spicules, when referring to these events as a class (\textit{e.g.,}  when discussing their formation mechanisms or models) and we use the term RBE when referring to their \ha\ component observed on the disk. Recently \cite{2019A&A...632A..96R} speculated that RBEs may also display return flows. \cite{Sekse_2013} utilized 0.88~s temporal cadence data to find that the RBE life time ranges from 5 to 60~s and their transverse velocities may reach up to 55\kms. \cite{1998SoPh..178...55W} described a new type of small-scale chromospheric events, which they called upflows and suggested that they may be fueling coronal heating. They were linked to magnetic reconnection between the existing network fields and opposite polarity inter-network fields and ephemeral regions. \cite{1998ApJ...504L.123C} attempted to associate the upflows with SUMMER UV explosive events and magnetic reconnection in QS areas. \cite{2000ApJ...545.1124L} reported that the majority of the upflow events show absorption only in the blue wing of the \ha\ line, which is similar to the RBEs \citep[e.g.,][]{2008ApJ...679L.167L}. 

Some of type II spicules appear to show twisting motions \citep[e.g.,][]{Tomczyk2007,torsional_osc} or doubling \citep{2008ASPC..397...27S}, and they have been used as tracers for Alfv{\'e}nic waves \citep{2007Sci...318.1574D}. Their identification in IRIS and AIA images and the increased line broadening suggests that they are heated to at least the transition region temperatures \cite[e.g.,][]{2007PASJ...59S.655D,2007Sci...318.1574D,2009ApJ...701L...1D,torsional_osc,2016ApJ...820..124H}. When observed in the \textit{Hinode} Ca II band, they show fading on time-scales of an order of tens of seconds \citep{2007Sci...318.1574D}. Type II spicules and RBEs are subject to various high frequency oscillations \citep[e.g.,][]{Okamoto_Bart,Sekse_2013}. Although it is not clear what drives type II spicules, two energy sources may be considered: leakage of p-mode waves into the chromosphere or release of magnetic energy either via release of magnetic tension \cite{2017Sci...356.1269M}, oscillatory reconnection \citep{2012ApJ...749...30M}, or magnetic reconnection \citep{2013ApJ...767...17Y,2015ApJ...799..219D,Sci2019}. \cite{Judge_2012} argued that spicules II could be warps in 2D sheet-like structures, while \cite{zhang12revision} questioned the existence of spicules II as a distinct class altogether.

The formation process of type II spicules is thought to affect the corona by generating shocks, flows, waves and currents, which can be linked to other phenomena such as the red-blue asymmetries observed in UV data as well as propagating coronal disturbances observed with the 17.1~nm and 19.3~nm SDO/AIA channels \citep[e.g.,][]{1998ApJ...504L.123C,2018ApJ...860..116M}. Their detailed physical cause and role in providing mass and energy to the corona remain largely unknown. The related difficulties in the interpretation of solar data mainly arise from the limited spatial resolution and complexity of the chromosphere \citep[e.g.,][]{0004-637X-749-2-136}. They appear in regions of seemingly unipolar magnetic fields often surrounding clusters of photospheric bright points \citep[e.g.,][]{2009ApJ...706L..80M}. \cite{2011Sci...331...55D} suggested that they may be a product of reconnection. \cite{2013ApJ...767...17Y} reported that the occurrence of packets of type II spicules is generally correlated with the appearance of new, mixed or unipolar fields in close proximity to network fields. These authors also suggested that emerging fields may introduce rapid reconfiguration of equilibrium in the pre-existing fields, which may further lead to both small scale (component) reconnection and high frequency MHD waves. Detection of kinked and/or inverse ``Y'' shaped RBEs further confirms this conclusion. Recently \cite{Sci2019} observed that RBEs appeared in the Goode Solar Telescope (GST) data within minutes after opposite-polarity magnetic flux appeared around a cluster of dominant polarity. 

\cite{2017ApJ...849L...7D} used 2.5D radiative MHD simulations to show that spicules can be driven by ambipolar diffusion resulting from ion-neutral interactions. \cite{2017Sci...356.1269M} further advanced the study and found that simulated spicules occur when magnetic tension is amplified and transported upward through interactions between ions and neutrals or ambipolar diffusion. The tension is impulsively released to drive flows, heat plasma (through ambipolar diffusion), and generate Alfv{\'e}nic waves. The magnetic tension is introduced in the system through new flux emergence that undergoes reconnection with the ambient fields. It is important to stress that the simulated spicules were not accelerated by the reconnection event. Nevertheless, none of the current models can explain all of the observed properties of type II spicules and RBEs including their omnipresence \citep{2007PASJ...59S.655D,2012ApJ...759...18P}, temperature, and the associated wave energy \citep[e.g.,][]{Liu2019}.

\section{Data} \label{sec:data}

On June 7 2019 GST \citep{2010ApJ...714L..31G, Cao2010} acquired QS data near the disk center at heliocentric-cartesian position (-115\asec, 135\asec) with the aid of an adaptive optics system. We only used data from Visible Imaging Spectrometer (VIS) that utilizes a Fabry-P\'{e}rot interferometer with a bandpass of 0.008~nm and the possibility to shift the bandpass by 0.2~nm around the \ha\ line center. The pixel scale was set to 0\asec.027 and the field of view (FOV) of the imager was 75\asec$\times$64\asec. RBEs display notable Doppler shift and line broadening and are best seen in the blue wing of the \ha\ spectral line. Generally speaking the number of RBEs seen in the FOV decreases with the increasing distance from the \ha\ line center. However, when observing closer to the line center the FOV of view is contaminated by overlying fibrils with their own flows. To acquire high cadence data we limited our choice of usable wavelengths to one spectral position (\ha-0.08~nm). We also acquired a short series of \ha-0.08 and \ha+0.08~nm pairs that allowed us to produce \ha 0.08~nm Doppler maps at a 5~s cadence. The original data was acquired in bursts of 25 frames each. All bursts were speckle reconstructed using \cite{kisip_code} technique to produce the final \ha-0.08~nm images with 2~s cadence, which were aligned and de-stretched to remove residual image distortion due to seeing and telescope jitter. The intensity of each image was adjusted to the average level of the data set.

\section{Results} \label{sec:results}

\begin{figure}[ht]
\epsscale{0.7}
\plottwo{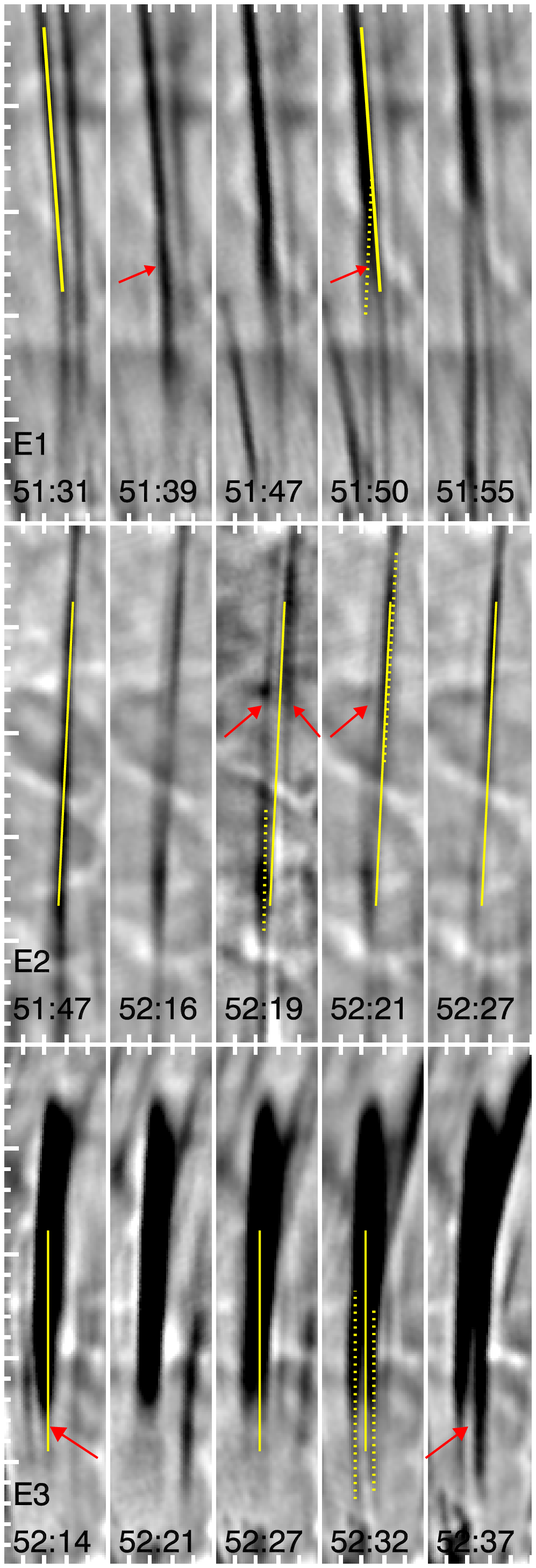}{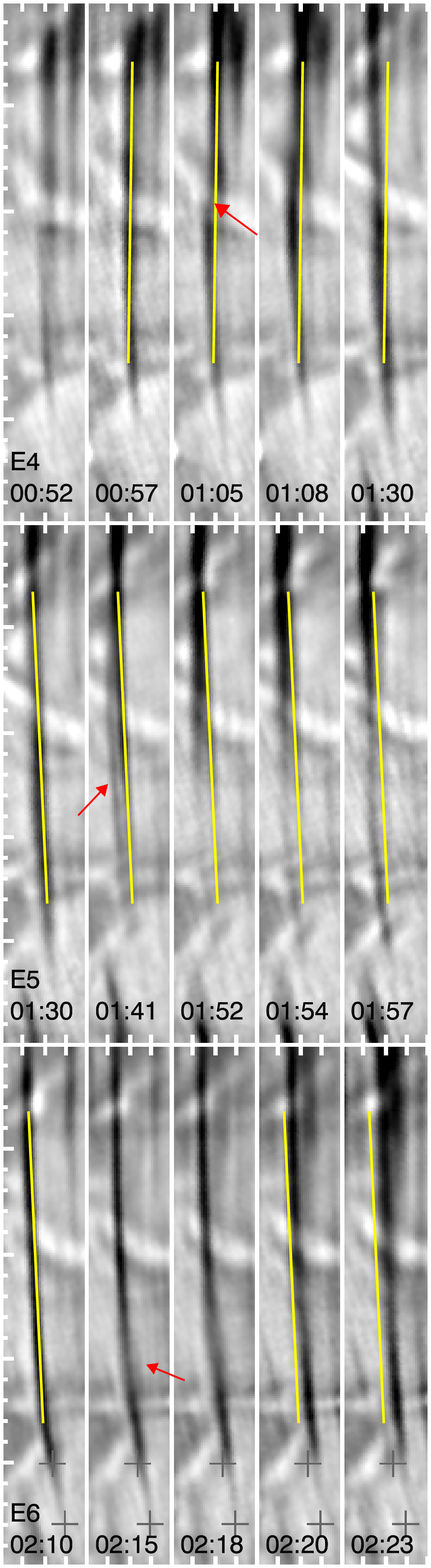}
\caption{Evolution of events 1--6 as seen in GST/VIS \ha-0.08~nm data. The over-plotted solid yellow lines indicate the initial position of the evolving RBE, while dashed lines highlight newly formed features. Arrows point toward various features that are discussed in the text. Note that the format of the time stamp at the bottom of each panel is MM:SS and the short tick marks separate 0.2~Mm (0\asec.275) intervals. Crosses in the E6 panels mark two initial positions of the RBE endpoint and are plotted to ease comparison. \label{e16}}
\end{figure}

Here we discuss data for 12 RBE events that showed very rapid and distinct evolutionary patterns. Figure \ref{e16} shows evolution of events 1 -- 6. Note that the bright lanes seen in the background are chains of bright points seen in the intergranular lanes. Event E1 lasted for about 30~s and started as broadening of the pre-existing feature (solid line in 51:31 panel and arrow in 51:39 panel) followed by a double RBE that appeared to be joined at the top thus forming the inverted ``Y'' pattern. The two ``legs'' that extend down from the location indicated by the red arrow in 51:50 panel (dashed lines) were developed during this process and they appear in \ha\ to be extending downward toward the photosphere. The feature widening was not accompanied by an injection of chromospheric plasma observable in the \ha\ line into the volume as is typical for a chromospheric jet. \cite{2012ApJ...752..108S} reported that the lower endpoints of Ca II 854.2~nm component of RBEs are located closer to the network, so that injection of cool plasma may still occur without being detectable in the \ha\ line. We also point out sudden appearance ($<8$~s) and equally rapid disappearance ($<5$~s) of an RBE feature (not related to E1) seen at the bottom of 51:39-51:55 panels of event E1. Other similar examples will be discussed further in the text. Event E3 (bottom row) is another example of RBE doubling although this feature was much broader, darker, and longer living. 

Event E2 displays a different type of evolution where an existing RBE (solid line in 51:47 panel) suddenly dimmed and became fuzzy in the midsection. In each of these panels, the solid line marks the initial position of the RBE as it was recorded at 51:47. Very faint dark strands then appeared on both sides of the RBE. To enhance the strands, we subtracted the background and the residual image in the 52:19 panel shows the enhanced features (arrows). The strands appeared to be moving away from the original RBE at a rate of about 25\kms, while the main feature remained stationary (solid line in 52:27).

Event E4 represents a ``fractured'' RBE formed after it suddenly darkened, and broke into two branches displaced in the horizontal direction (arrow in 01:05 panel). Note that the lower part at that moment is no longer co-spatial with the initial position (yellow line). In a matter of seconds the upper branch disappeared, while the lower part extended upward thus forming a new RBE feature displaced by about 0\asec.3 from the initial position over a period of time of about 5-7~s resulting in the displacement rate of about 30-50\kms.

Events E5 and E6 represent other cases of dimmed and diffuse evolution of RBEs. Similar to the E2 event, a faint side line appeared on one side of E5 (arrow in 01:41 panel), and it soon developed into a regular RBE feature. Event E6 developed a very compact, well defined, oval-shaped diffuse region (arrow in 02:15 panel) of about 0\asec.2~Mm wide and 1~Mm long. The lower branch of the RBE (below the diffuse region) showed later displacement by about 0.2~Mm. It is not clear whether the upper part was displaced as well or a new feature formed at that location. Similarly to the previous cases the displacement rate was estimated to be about 30-50\kms.

\begin{figure}[ht]
\epsscale{1.2}
\plotone{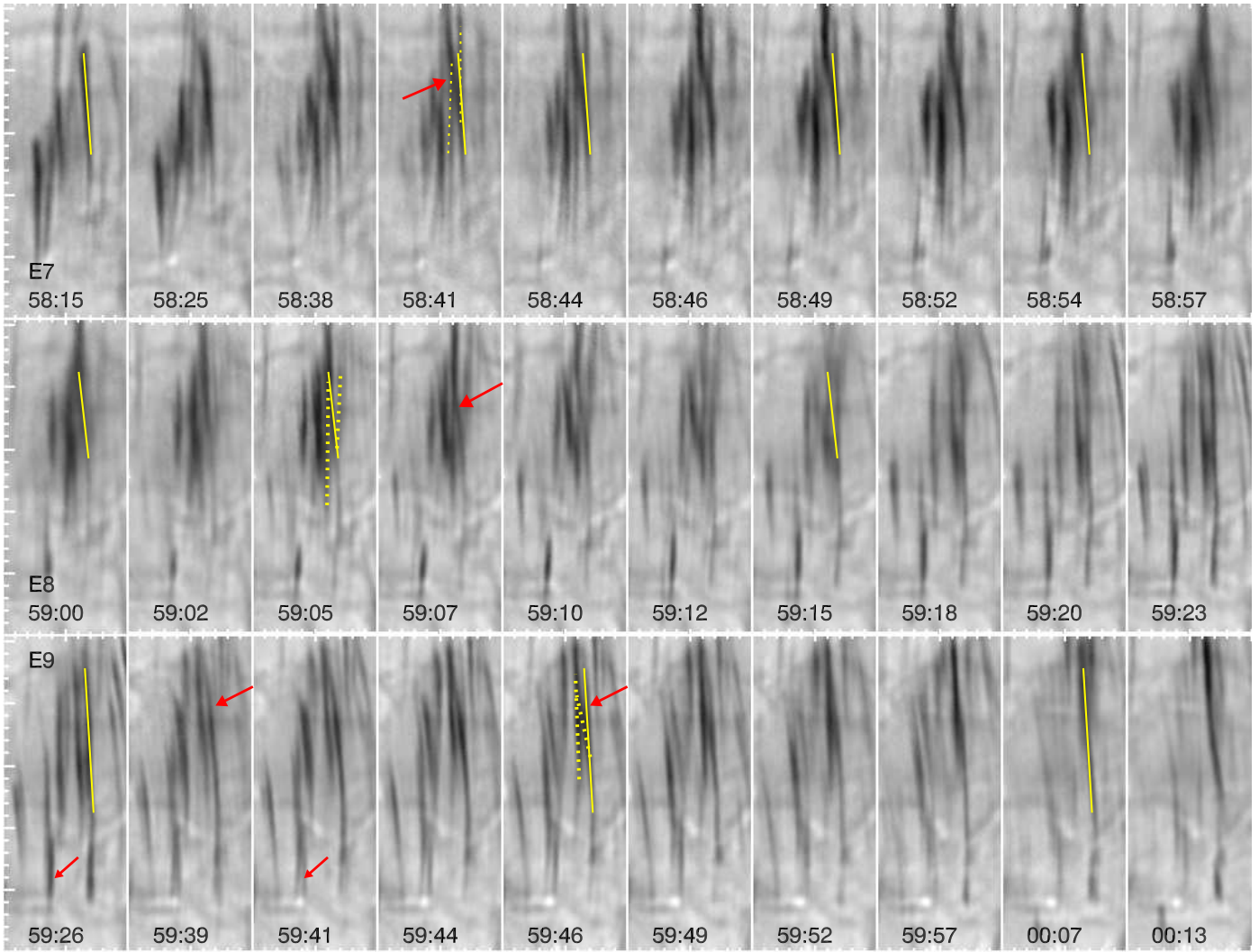}
\caption{The same as in Fig. \ref{e16} but for events 7--9. 
\label{e79}}
\end{figure}

In Figure \ref{e79} we show the same QS area with three ``N'' pattern features. The arrow in 58:41 panel (top row) shows an ``N'' pattern consisting of two vertical and a slanted RBE streaks. This pattern typically starts from one slanted feature (E7, the yellow line in 58:15 panel) that rapidly evolves first into an x-configuration and then into an ``N'' configuration as the new features form and separate. The ``N'' pattern then changes into what appears to be a ``kinked'' spicule (58:52), which gradually straightens (58:57 panel). However, it is not clear if that ``kinked'' RBE was indeed one feature or a composition of several smaller components.

Apart from the most prominent RBE transformation cases that we describe here, there also are other sudden \textit{in situ} appearances of RBEs. Comparison of the 58:15 and 58:25 panels of event E7 shows that several dark striations have appeared within a 10~s time interval on the left side of the the yellow line shown in the 58:15 panel. Similarly, several new features appeared in the same area between 58:38 and 58:49 panels. A new, ``Y'' shaped feature is visible in the center of the 59:44 panel of event E9 that was not present yet in the 59:30 panel.

Evolution of event E8 is similar to E7 with the only distinction that the ``N'' pattern (arrow in 59:07 panel) evolved into two a well defined double RBE feature with component separation of about 0.3~Mm. Event E9 is another example of the ``N'' pattern that involved an RBE with swaying motion (compare 59:26 - 59:41 panels). We should note that the swaying RBE was formed only about 10~s prior to the E8 event. 

We also note an RBE event that started as a single streak (red arrow in the lower left corner of the 59:26 panel of Figure \ref{e79}) and within 15~s evolved into a double feature with its lower endpoints at the opposite sides of a developing brightening and the upper endpoints still joined at a distance of about 1~Mm from the brightening (panel 59:39). Note that the brightening was not yet present in the 59:26 panel. This conjoined feature was observed to completely separate 5~s later (panel 59:44).

\begin{figure}[ht]
\epsscale{1.2}
\plotone{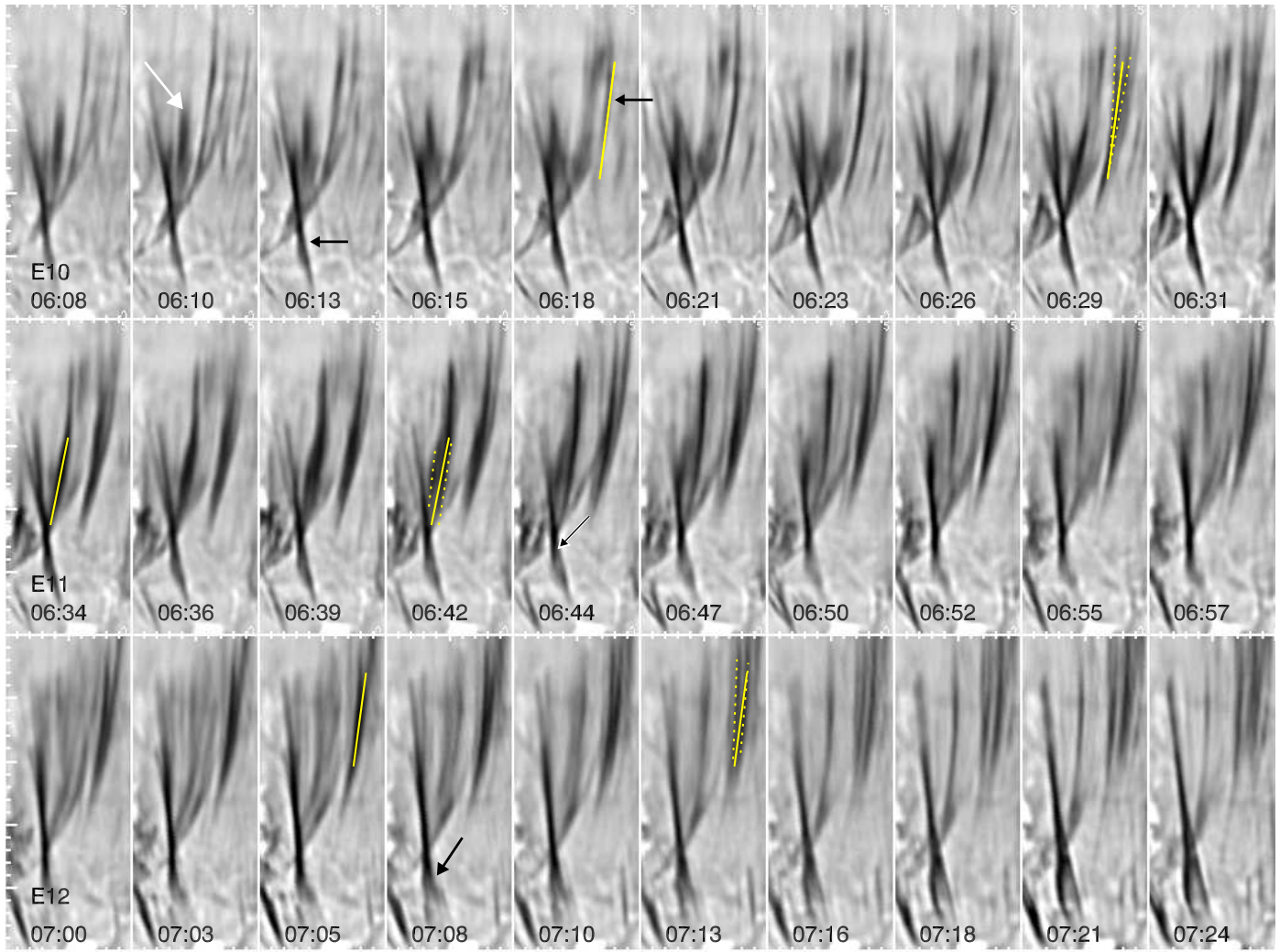}
\caption{The same as in Fig. \ref{e16} but for events 10 -- 12.
\label{e101}}
\end{figure}

Finally, in Figure \ref{e101} we show a ``V'' splitting pattern (E10) where a pre-existing RBE (black arrow and yellow solid line in 06:18 panel) evolves into a double feature that appear to be joined at their roots (opposite of the inverse ``Y'' splitting discussed above). The transition from a single to a double feature occurred within a 6~s interval (compare 06:23 and 06:29 panels). We should note that one of the double spicules split again following the ``V'' pattern (E12, 07:00 panel) and the new features later separated into nearly parallel structures (07:16 panel). The E11 event showed a parallel splitting or doubling. The original feature seen in 06:34 panel (solid yellow line) widened and 6~s later it is already seen split (06:42 panel). The components were moving away from each other and faded in about 10~s.

The last event in our series, E12 (arrows in 06:13, 06:44, and 07:08 panels), is quite different from the previous ones. It resembles an inverted ``Y'' \citep[or anemone, \textit{e.g.,}][]{Shibata1591} jet. It started before 06:08 and first appeared as a regular RBE. However, it had a much longer lifetime and was extending upward gradually developing into a typical reconnection jet with clear footpoint separation of about 500~km (07:24 panel).

\begin{figure}[ht]
\epsscale{0.5}
\plotone{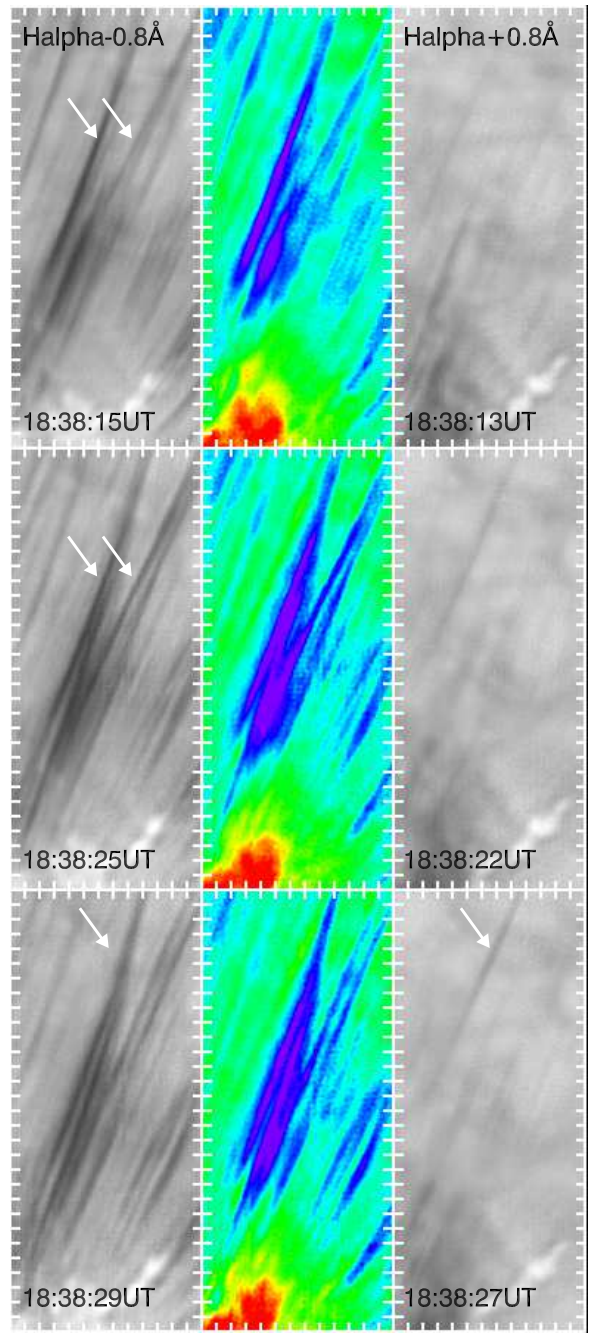}
\caption{\ha-0.08~nm (left), \ha+0.08~nm (right) and the corresponding \ha Doppler image (middle). The tick marks separate 0.2~Mm (0\asec.275) intervals. \label{doppler}}
\end{figure}

While these evolutionary patterns may, generally speaking, result from overlapping of multiple features, we argue that the cases considered here are not contaminated by that effect. Our arguments are primarily based on the fact that the evolving features showed structural changes (widening, darkening, or disappearance) before splitting into two parts or doubling. In most cases no features other than the evolving one were present at the scene at the beginning of the RBE transformation. To demonstrate the difference, we point to the fan-shaped system of loops seen in Figure \ref{e101} (arrow in 06:10 panel) that showed rapid transverse displacement. The system was observed moving in the background behind the jet (arrow 06:13 panel) without any detectable interaction with it or other stationary features. In contrast, quite often a complex ``N'' patterns develop from a single \textit{stationary} feature initially present in the FOV (e.g., E7 in Figure \ref{e79}). 

Figure \ref{doppler} shows \ha$\pm$0.08~nm images (left and right columns) and the corresponding Doppler image (middle) of evolving RBEs at three instances. The Dopplergrams of two splitting RBEs show that prior to and during the event, the evolving features appeared blue-shifted in the GST/VIS instrument. However, we did not observe detailed spectra of the event, which is needed to perform an accurate analysis of the line of sight speeds associated with these RBE transformations. 

\section{Summary and Discussion}

We inspected a 30 min long data set and found that the RBE transformations described above are very frequent and ubiquitous. Although each new individual feature followed its own unique evolutionary path, they often exhibit group behaviour \citep{2011Natur.475..477M} when several strands follow coherent swaying motions. 
We thus summarize our findings as follows: i) very frequently RBEs suddenly appear \textit{in situ} as non-extending plasma structures without prior \ha\ features and clear evidence of \ha\ emitting material being injected from below; ii) they rapidly evolve on time scales of the order of 1~s; iii) their evolution includes inverted ``Y’’, ``V’’, ``N'', or parallel splitting (doubling) as well as sudden formation of a diffuse region followed by branching; and iv) the same feature may undergo several splitting episodes within about 1 min time interval. 

\cite{Sekse_2013} interpreted the sudden appearance of RBEs over their full length as a combination of field-aligned flows, transverse swaying, and torsional motions \citep[e.g.,][]{torsional_osc}. \cite{2011Sci...331...55D} noted that the appearance of AIA type II spicules is delayed relative to the corresponding \ha\ feature \cite[also ][]{2014ApJ...792L..15P}. At the same time \cite{Sekse_2013} reported that Ca II 854.2~nm RBEs may appear earlier than their \ha\ component. \cite{2015ApJ...806..170S} reported that when Ca II spicules fade from the pass-band they continue to be visible in other ``hotter'' spectral lines. These data suggest that spicules are heated during their life time. This also implies that the \ha\ component of type II spicules (RBEs), considered here may appear in the FOV before any other hotter component as a result of heating of cool plasma detected in Ca II 854.2~nm \citep{Sekse_2013}. However, their sudden \textit{in situ} appearance in the \ha-0.08~nm spectral window via transformation of existing features does not seem to be consistent with any of the models that interprets these events as jets originating in the lower chromosphere \citep[e.g.,][]{2018ApJ...856..176G}. To the contrary, in some cases we seem to observe the opposite effect: during their transformation process some \ha-0.08~nm RBEs are seen extending downward toward the photosphere.

Recently \cite{Cho_2019} reported on a new type of faint chromospheric jets detected at a limb in Ca II \textit{Hinode} images. These jets exhibit an average speed of about 132\kms and an average lifetime of 20~s, ranging from 11 to 36~s. These new limb jets could be related to the short lived RBEs discussed here. These jets appear to originate 2-3~Mm above the limb, which roughly places them at the top of \ha\ RBEs. 

\cite{Sci2019} identified magnetic flux cancellations at the footpoint of some analyzed RBEs. Two events considered in the present study do exhibit morphological features and evolutionary pattern consistent with the flux cancellation scenario. On the other hand, new flux emergence may trigger enhanced spicule activity observed as bursts of RBEs \citep{2013ApJ...767...17Y,Sci2019}. These may suggest that in addition to directly driving the RBEs, flux emergence may also cause both re-structuring of coronal fields and wave generation \citep[e.g.,][]{isobe2008,2011ApJ...736....3V,2017NatSR...743147S}, which in turn may be responsible for the rapid RBE transformations reported here. Furthermore, waves are also generated by vigorous turbulent flows \citep[e.g.,][]{2018ApJ...866...73A,Liu2019}. The MHD waves experience reflection near the TR and propagate downward \citep{Okamoto_Bart}. All these may create complex and dynamic interactions when coronal fields withing a flux tube attempt to counter the de-stabilizing driving force and injection of energy \citep{Ascw2018} via multiple small-scale reconnection events occurring throughout the volume. Additionally, the short lifetime and sudden appearance of these events may be a manifestation of the sheet-like structures as conjectured by \cite{Judge_2012}.

The events presented here resemble the well-known idea of inter-twined and tangled magnetic field lines \citep{Parker1989}. The criss-crossing patterns formed by RBEs are also somewhat similar to the synthetic emission of UV strands heated via multiple reconnections of tangled field lines within a flux tube \citep[see Figures 5-7 in][]{2017ApJ...837..108P}. However, while it is tempting to interpret the RBE transformations as driven by component magnetic reconnection, we should be mindful that according to \cite{2017ApJ...837..108P} the appearance of energy release regions associated with reconnection in a braided magnetic structure may be wavelength and geometry dependent. The data presented here were acquired in only one, very narrow spectral range and multi-wavelength data sets need to be analyzed to learn about the temperature and flow patterns associated with the locations where RBEs transformations occur and to offer a plausible explanation.

\acknowledgments

BBSO operation is supported by NJIT and US NSF AGS-1821294 grants. GST operation is partly supported by the Korea Astronomy and Space Science Institute (KASI), Seoul National University, the Key Laboratory of Solar Activities of Chinese Academy of Sciences (CAS), and the Operation, Maintenance and Upgrading Fund of CAS for Astronomical Telescopes and Facility Instruments. V.Y. acknowledges support from NSF AST-1614457, AFOSR FA9550-19-1-0040 and NASA 80NSSC17K0016, 80NSSC19K0257, and 80NSSC20K0025 grants. X.Y. was supported by NSF 1821294, 1614457, and National Science Foundation of China 11729301 grants. K-S.C. acknowledges support from KASI under the R\&D program ``Development of a Solar Coronagraph on International Space Station'' (project No. 2020-1-850-02) supervised by the Ministry of Science and ICT.

\vspace{5mm}
\facility{GST}



\end{document}